\documentstyle[preprint,aps]{revtex}
\begin{document}
\draft
\title{Coherent manipulation of motional states of trapped ions}
\author{Hao-Sheng Zeng$^{1,2}$\thanks{%
E-mail adress: hszeng@mail.hunnu.edu.cn}, Le-Man Kuang$^{1}$ and Ke-Lin Gao$%
^{2}$}
\address{$^{1}$Department of Physics, Hunan Normal University, Hunan\\
410081, People's Republic of China\\
$^{2}$Laboratory of Magnetic Resonance and Atomic and Molecular\\
Physics, Wuhan Institute of Physics and Mathematics, Chinese Academy of\\
Science, Wuhan 430071, People's Republic of China }
\maketitle

\begin{abstract}
The beam splitter and phase shifter, which are the key elements in the
experiments of light interference, are realized in the motion of trapped
ions. Some applications, such as the creation of quantum motional states and
the realization of Mach-Zehnder interferometer, are illustrated. Several
detection methods of motional states used in the interferometer are also
discussed.
\end{abstract}

\pacs{Keywords: Quantum motional states, beam splitter, phase shifter.}

\vskip 1cm

\narrowtext

{\bf 1. Introduction}

Beam splitter is a very important element in quantum optical experiments
which can realize the splitting and mixing of the input light fields to
produce correlated or entangled output states. A number of authors have
considered the behavior of the quantum-mechanical beam splitter in the past
few years [1-6]. In nearly every interferometric apparatus such as
Mach-Zehnder[1, 7-9], Fabry-Perot[8] and Michelson-type[10] interferometers,
the beam splitter plays a central role. Both passive and active
interferometers [10-12] that have been touted to revolutionize
ultrasensitive measurements like gravity-wave detection, sensing of tiny
rotation rates, etc. use beam splitters. In addition, phase shifter, which
may be realized by changing the length of an optical path or by employing
the optical Kerr effect [10, 13], is also an important ingredient in most
optical interference experiments.

On the other hand, the developments in the cooling and trapping ions[14]
have opened a new research for both atomic physics and quantum optics. When
ions trapped in an electromagnetic harmonic potential, the quantized
vibratic motions play a role of boson modes. And the interaction between
ions and laser fields leads the coupling between the external and internal
degrees of freedom of the trapped ions[15-16]. As the coupling between the
vibratic modes and the external environment is extremely weak, dissipative
effects, which are inevitable from cavity damping in the optical regime, can
be significantly suppressed. This unique feature thus can be used to realize
some cavity QED experiments without using an optical cavity. Stimulated by
these ideas, many schemes have been proposed for generating nonclassical
motional states of trapped ions, such as Fock states [16], squeezed state
[17-19], even and odd coherent states [20-23] and their various kinds of
superpositions[24-29]. These nonclassical motional states may be used to
perform quantum computation[30-32], test quantum-mechanical basic
principals[32] and study quantum decoherence effects[33]. Up to now,
motional Fock states, squeezed states, coherent states [34-35] and Schr$%
\ddot{o}$dinger cat [36] for the c.m. mode of a single trapped ion have been
produced and observed experimentally. The reconstruction of motional states
have also been realized in experiments[37].

However, the ability that people coherently manipulate motional states is
much lower than that people manipulate light fields. In order to further
enrich and develop the content of this portion, we here present a method to
implement the beam splitter and phase shifter in the motion of trapped ions,
and then use them to coherently produce some typical quantum motional states
and to implement the famous Mach-Zehnder interferometer in the vibrational
motion of trapped ions. Some of the results or ideas may also be used
directly to implement other vibrational interferential experiments, such as
the homodyne and Michelson-type interferometer. The construction of this
paper is arranged as follows: In section 2, we first describe how to
construct the two pivotal elements for implementing a Mach-Zehnder
interferometer--- the two-mode beam splitter and phase shifter for the
motional states of c.m. and breathing modes in two trapped ions, where the
later is mainly based on the dynamical evolution of laser-drived trapped
ions first introduced by D'Helon and Milburn[38]. In section 3, we then
illustrate their applications in the generation of quantum motional states.
And in section 4, we further use them to implement a Mach-Zehnder
interferometer for the vibrational motion of trapped ions, where three
detection ways of motional states are provided. The first two ways are
mainly based on the idea of the reconstruction of motional states[34, 39],
and the third one based on the direct measurement of the mean value of a
motional operator[40]. A conclusion is given in section 5.

{\bf 2. Beam splitter transformation and phase shift in the motion of
trapped ions}

Beam splitter and phase shifter are two key elements in quantum optical
experiments. Their realizations in the motion of trapped ions will allow us
to implement the similar interferential experiments in trapped-ion systems
without the use of optical elements. Though some papers have already
involved this problem[41-42], the general constructions of beam splitter and
phase shifter in trapped-ion systems need further developments. Here we
provide a method to realize this two important elements in the motion of two
trapped ions.

Following the experimental implementation of reference [43], we consider a
two-level system which is formed by two $^{9}$Be$^{+}$ ions of mass $m$
trapped in a linear trap. The two spectrally resolved ground-state hyperfine
levels are defined as the ground and excited states of our problem: $\left|
F=2,m_{F}=-2\right\rangle \equiv \left| g\right\rangle $ and $\left|
F=1,m_{F}=-1\right\rangle \equiv \left| e\right\rangle $, which form the
so-called internal degrees of freedom. Assume that the two ions are strongly
bounded in the $y$ and $z$ directions, but weakly bounded in a harmonic
potential in the $x$ direction and oscillate around their equilibrium
positions: $x_{10}=-d/2$, $x_{20}=d/2$. We denote by $\widehat{X}=(\widehat{x%
}_{2}+\widehat{x}_{1})/2$, $\widehat{x}=(\widehat{x}_{2}-\widehat{x}_{1})/2$
the c.m. and breathing mode operators respectively which constitute the
studied external vibratic motional degrees of freedom. Now we let only one
of the ions, say the first ion, to be illuminated simultaneously by two
classical standing wave lasers with the positive frequency parts $%
E_{I}^{(+)}=E_{0I}\cos (k_{I}x+\varphi _{I})e^{-i\omega _{I}t-i\phi _{I}}$
and $E_{II}^{(+)}=E_{0II}\cos (k_{II}x+\varphi _{II})e^{-i\omega
_{II}t-i\phi _{II}}$. Assume that the two lasers have the same amplitudes $%
E_{0I}=E_{0II}=E_{0}$ and effective wavevectors $k_{I}=k_{II}=k$, but
different frequencies $\omega _{I}$ and $\omega _{II}$. Experimentally, the
two standing wave lasers with needed properties can be implemented by
employing two-photon stimulated-Raman transitions[18, 31]. Of course, in
order to let a laser illuminate selectively only ion $1$ without lighting
ion $2$, the laser propagation should form an appropriate angle with $x$
axis[44] and the projective wave in the direction of $x$ axis plays the role
of coupling the internal and the wanted external degrees of freedom.
Assuming that the ion is located at the anti-nodes of both the two standing
waves, i.e. $\varphi _{I}=\varphi _{II}=0$, then the Hamiltonian for this
system can be written as,

\begin{equation}
H=H_0+H_{int},
\end{equation}

\begin{equation}
H_{0}=\mu a^{+}a+\nu b^{+}b+\omega _{0}\sigma _{z1}/2,
\end{equation}

\begin{equation}
H_{int}=\frac{\Omega }{2}\sigma _{+1}[(e^{-i\omega _{I}t-i\phi
_{I}}+e^{-i\omega _{II}t-i\phi _{II}})\cos (k\widehat{x}_{1})+h.c.]
\end{equation}
where $\mu $ ($\nu $) and $a$ ($b$) are the frequency and annihilation
operator of the c.m. mode (breathing mode), $\omega _{0}$ the energy
difference between ground state $\left| g\right\rangle $ and excited state $%
\left| e\right\rangle $ of each ion, $\sigma _{z1}$ and $\sigma _{+1}$ the
usual Pauli operators describing the internal degrees of freedom for ion $1$%
, $\Omega $ the Rabi frequency for the ion-laser coupling which is assumed
to be equal for the two standing wave lasers.

We select the two standing waves to excite resonantly both the upper and
lower sidebands of c.m. and breathing modes, i.e. the frequencies of the two
lasers satisfying

\begin{equation}
\omega _{I}=\omega _{0}-(\mu -\nu ),\quad \omega _{II}=\omega _{0}+(\mu -\nu
).
\end{equation}
Under the condition that both c.m. and breathing modes are cooled to
Lamb-Dicke regime, we can expand the Hamiltonian up to the order terms of $%
\eta ^{2}$ or $\eta _{r}^{2}$ which represent the lowest couplings between
internal and external degrees of freedom. Transforming the above Hamiltonian
to the interaction picture with respect to $H_{0}$, we have 
\begin{equation}
H_{int}^{1}=\Omega \eta \eta _{r}(a^{+}b+ab^{+})\sigma _{x1}
\end{equation}
for one kind of laser-phase selection of $\phi _{I}=\phi _{II}=0$ and

\begin{equation}
H_{int}^{2}=-i\Omega \eta \eta _{r}(a^{+}b-ab^{+})\sigma _{x1}
\end{equation}
for another kind of laser-phase selection of $\phi _{I}=-\phi _{II}=-\pi /2$%
, where $\eta =k\sqrt{1/4m\mu }$ and $\eta _{r}=k\sqrt{1/4m\nu }$ are the
Lamb-Dicke parameters corresponding to the c.m. and breathing modes, $\sigma
_{x1}$ the $x$ component of Pauli operator for ion $1$. The exact
propagators for the above Hamiltonians are

\begin{equation}
U_{1}=\exp \{-i\Omega \eta \eta _{r}t(a^{+}b+ab^{+})\sigma _{x1}\}
\end{equation}
for the first kind of laser-phase selection and

\begin{equation}
U_{2}=\exp \{-\Omega \eta \eta _{r}t(a^{+}b-ab^{+})\sigma _{x1}\}
\end{equation}
for the later kind of laser-phase selection. Note that the interactions of $%
U_{1}$ and $U_{2}$ occur only between the internal states of ion $1$ and
motional states of the system. Now we prepare the initial state of the whole
system as

\begin{equation}
\left| \psi \right\rangle =\left| \psi \right\rangle _{1}\otimes \left| \psi
\right\rangle _{2}\otimes \left| \psi \right\rangle _{c,r}
\end{equation}
with$\left| \psi \right\rangle _{1}=(\left| e\right\rangle _{1}+\left|
g\right\rangle _{1})/\sqrt{2}$ and $\left| \psi \right\rangle _{2}$ (which
may be a arbitrary pure state) the internal states of the two ions and $%
\left| \psi \right\rangle _{c,r}$ the initial motional state of c.m. and
breathing modes. By performing unitary propagators $U_{1}$ or $U_{2}$ on
this initial state respectively, we find that both the internal states of
the two ions remain unchanged but the motional states experiences a
beam-splitter transformation: 
\begin{equation}
B_{1}(\theta )\left| \psi \right\rangle _{c,r}=e^{-i\theta J_{x}}\left| \psi
\right\rangle _{c,r},
\end{equation}
or

\begin{equation}
B_{2}(\theta )\left| \psi \right\rangle _{c,r}=e^{-i\theta J_{y}}\left| \psi
\right\rangle _{c,r}
\end{equation}
respectively, where $B_{1}(\theta )=\exp (-i\theta J_{x})$ and $B_{2}(\theta
)=\exp (-i\theta J_{y})$ with $\theta =2\Omega \eta \eta _{r}t$ being a real
angle. The Schwinger angular-momentum operators [45] are defined as

\begin{eqnarray*}
J_{x} &=&(a^{+}b+ab^{+})/2, \\
J_{y} &=&(a^{+}b-ab^{+})/2i, \\
J_{z} &=&(a^{+}a-b^{+}b)/2,
\end{eqnarray*}
which satisfy the usual commutation relations for the Lie algebra of SU(2): $%
[J_{x},J_{y}]=iJ_{z}$, $[J_{y},J_{z}]=iJ_{x}$ and $[J_{z},J_{x}]=iJ_{y}$.
The transformations $B_{1}(\theta )$ and $B_{2}(\theta )$ represent
rotations by angle $\theta $ about $-x$ and $-y$ axes in the
angular-momentum space respectively, which would be distinguished as $\pi
/2- $ and $\pi -$ coupling beam splitters[1] for c.m. and breathing modes.
When $\theta =\pi /2$, they are referred to as 50/50 beam splitters.

The phase shift transformation for motional states of one trapped ion may be
realized by changing the trap frequency for a fixed time[46]. Here we
construct the phase shifters for c.m. and breathing modes of two trapped
ions using the dynamical evolution of laser-drived trapped ions first
introduced by D'Helon and Milburn[38]. For convenience in the next sections,
we consider an interaction of a classical standing wave with ion $2$ in the
above two-ion system. Assume that the ion is placed at the node of the
standing wave and the detuning $\Delta =\omega _{0}-\omega _{L}$ is set to
excite only the c.m. motional degrees of freedom but far away from the
corresponding sideband resonance ($|\Delta |\gg \mu $). In Lamb-Dicke limit
and for interaction times much greater than the vibrational period of c.m.
mode, this interaction produces the following conditional phase shift[38,42]

\begin{equation}
U_{3}=\exp [-i\chi ta^{+}a(\sigma _{z2}+1/2)],
\end{equation}
where $\chi =\eta ^{2}\Omega ^{2}/(2\Delta )$ with $\eta $ the Lamb-Dicke
parameter of c.m. mode and $\Omega $ the Rabi frequency for the ion-laser
coupling. Pauli operator $\sigma _{z2}$ describe the internal degrees of
freedom for ion $2$, and $t$ is the interaction time. The conditional phase
shift $U_{3}$ occurs only between the internal state of ion $2$ and the c.m.
motional state which can be used to realize a phase shift transformation on
c.m. mode. For example, by preparing the initial state of the involved
system as $\left| g\right\rangle _{2}\otimes \left| \psi \right\rangle _{c}$
with $\left| g\right\rangle _{2}$ the ground state of ion $2$ and $\left|
\psi \right\rangle _{c}$ the vibrational state of c.m. mode, then the action
of $U_{3}$ on this state produces a phase shift on c.m. mode:

\begin{equation}
P_{c}(\varphi )\left| \psi \right\rangle _{c}=\exp (i\varphi a^{+}a)\left|
\psi \right\rangle _{c}
\end{equation}
without changing the internal state of ion $2$. The phase-shift angle $%
\varphi =\chi t/2$ may be adjusted at will by means of altering the
interaction time.

The phase shift on breathing mode may also be realized in a similar way. In
the above process, if we use the dispersive standing wave to excite
non-resonantly the breathing mode instead of c.m. mode, in the similar
conditions one has

\begin{equation}
U_{4}=\exp [-i\chi _{r}tb^{+}b(\sigma _{z2}+1/2)],
\end{equation}
where $\chi _{r}=\eta _{r}^{2}\Omega _{r}^{2}/(2\Delta _{r})$ with $\eta
_{r}=3^{-1/4}\eta $ the Lamb-Dicke parameter for breathing mode, $\Omega
_{r} $ the Rabi frequency and $\Delta _{r}=\omega _{0}-\omega _{L}$ the
detuning frequency which satisfies $|\Delta _{r}|\gg \nu $ this time. By
preparing the initial state of the involved system as $\left| g\right\rangle
_{2}\otimes \left| \psi \right\rangle _{r}$ with $\left| \psi \right\rangle
_{r}$ the vibrational state of breathing mode, then, without changing the
internal state of ion $2$, the unitary propagator $U_{4}$ produces a phase
shift for breathing mode:

\begin{equation}
P_{r}(\varphi _{r})=\exp (i\varphi _{r}b^{+}b)
\end{equation}
with the phase-shift angle $\varphi _{r}=\chi _{r}t/2$.

It is worthwhile to point out that the similar two-mode beam splitter
transformation of eq.(11) can also be realized in the motion of one ion
trapped in a two-dimensional isotropic harmonic potential trap[41]. The
merit to use two ions instead of one ion is that it can realizes
simultaneously the two kinds of beam-splitter transformations of
eqs.(10)-(11) by simply adjusting the initial phases of the applied lasers.
Further, combining these two kinds of beam-splitter transformations with the
phase shifts given by eqs.(13) and (15), we can construct the most general
two-mode beam-splitter transformation[6]. The constructions of the above
beam splitter and phase shifter on motional states are simple, which employ
at most two standing waves to interact with one ion. Thus, it may be hopeful
to demonstrate in current or future experimental conditions.

{\bf 3. Creation of quantum motional states using beam splitter and phase
shifter}

As they played roles in the quantum optics, the beam-splitter and phase
shifter for motional states constituted in the last section may also possess
underlying applications. In this and next sections, we will show this
compendiously through concrete examples. In this section, let us first
describe the applications in producing quantum motional states. The first
kind of important motional states to be produced are the entangled number
states. By preparing the initial motional state in eq.(9) as $\left| \psi
\right\rangle _{c,r}=\left| 1\right\rangle _{c}\otimes \left| 0\right\rangle
_{r}\equiv \left| 1,0\right\rangle _{c,r}$ with $\left| 1\right\rangle _{c}$
and $\left| 0\right\rangle _{r}$ the Fock states of c.m. and breathing modes
respectively, and performing $U_{1}$ or $U_{2}$ on it, we obtain the output
states for motional states as

\begin{equation}
B_{1}(\theta )\left| 1,0\right\rangle _{c,r}=\cos (\theta /2)\left|
1,0\right\rangle _{c,r}-i\sin (\theta /2)\left| 0,1\right\rangle _{c,r},
\end{equation}

\begin{equation}
B_{2}(\theta )\left| 1,0\right\rangle _{c,r}=\cos (\theta /2)\left|
1,0\right\rangle _{c,r}+\sin (\theta /2)\left| 0,1\right\rangle _{c,r}
\end{equation}
with $\theta =2\Omega \eta \eta _{r}t$ which can be adjusted by altering the
interaction time $t$. When $\theta =\pi /2$, we obtain the maximally
entangled single-phonon number states $\left( \left| 1,0\right\rangle
_{c,r}-i\left| 0,1\right\rangle _{c,r}\right) /\sqrt{2}$ and $\left( \left|
1,0\right\rangle _{c,r}+\left| 0,1\right\rangle _{c,r}\right) /\sqrt{2}$.
Instead, if we prepare the initial motional state as $\left| \psi
\right\rangle _{c,r}=\left| 1,1\right\rangle _{c,r}$, then the output states
are

\begin{equation}
B_{1}(\theta )\left| 1,1\right\rangle _{c,r}=\cos \theta \left|
1,1\right\rangle _{c,r}-i\sin \theta \left( \left| 2,0\right\rangle
_{c,r}+\left| 0,2\right\rangle _{c,r}\right) /\sqrt{2},
\end{equation}

\begin{equation}
B_{2}(\theta )\left| 1,1\right\rangle _{c,r}=\cos \theta \left|
1,1\right\rangle _{c,r}-\sin \theta \left( \left| 2,0\right\rangle
_{c,r}-\left| 0,2\right\rangle _{c,r}\right) /\sqrt{2}.
\end{equation}
For $\theta =\pi /2$, the maximally entangled two-phonon number states $%
\left( \left| 2,0\right\rangle _{c,r}\pm \left| 0,2\right\rangle
_{c,r}\right) /\sqrt{2}$ have been created. These maximally entangled number
states play an important role in quantum measurement theory. Similar
motional states have been produced for one ion trapped in a two-dimensional
isotropic ion trap[41]. The initial motional number states used above have
been already prepared experimentally by simply applying a sequence of Rabi $%
\pi $ pulses of laser radiation on the blue sideband, red sideband or
carrier [34]. Therefore the method for producing above entangled phonon
number states is practical. Once the above beam splitter has been realized
in experiment, then the creation of the discussed motional states will
become truth.

The second kind of important motional states to be produced are the
entangled coherent states. According to the formula for beam splitter
transformation

\begin{equation}
B_{2}(\theta )\left| \alpha \right\rangle _{c}\left| \beta \right\rangle
_{r}=\left| \alpha \cos (\frac{\theta }{2})-\beta \sin (\frac{\theta }{2}%
)\right\rangle _{c}\otimes \left| \alpha \sin (\frac{\theta }{2})+\beta \cos
(\frac{\theta }{2})\right\rangle _{r}
\end{equation}
with $\left| \alpha \right\rangle $ and $\left| \beta \right\rangle $ are
motional coherent states, if we let $\left| \psi \right\rangle _{c,r}=N_{\pm
}[\left| \alpha \right\rangle _{c}\pm \left| -\alpha \right\rangle
_{c}]\otimes \left| 0\right\rangle _{r}$ with $N_{\pm }=[2(1\pm \exp
(-2|\alpha |^{2})]^{-1/2}$ the normalized constants, then equation (11)
leads to a pair of entangled odd and even coherent states: 
\begin{equation}
\left| \widetilde{\alpha };\widetilde{\beta }\right\rangle _{\pm }=N_{\pm
}[\left| \widetilde{\alpha }\right\rangle _{c}\left| \widetilde{\beta }%
\right\rangle _{r}\pm \left| -\widetilde{\alpha }\right\rangle _{c}\left| -%
\widetilde{\beta }\right\rangle _{r}]
\end{equation}
with $\widetilde{\alpha }=\alpha \cos (\theta /2)$ and $\widetilde{\beta }%
=\alpha \sin (\theta /2)$. For a 50/50 beam splitter, i.e. $\theta =\pi /2$,
one has $\widetilde{\alpha }=\widetilde{\beta }=\alpha /\sqrt{2}$ and
eq.(21) reduces to

\begin{equation}
\left| \widetilde{\alpha };\widetilde{\alpha }\right\rangle _{\pm }=N_{\pm
}[\left| \widetilde{\alpha }\right\rangle _{c}\left| \widetilde{\alpha }%
\right\rangle _{r}\pm \left| -\widetilde{\alpha }\right\rangle _{c}\left| -%
\widetilde{\alpha }\right\rangle _{r}].
\end{equation}
The preparing methods for these types of entangled coherent states in the
motion of one trapped ion have been provided [22-23, 26]. In the limit of $%
\widetilde{\alpha }\rightarrow \infty $ and $\widetilde{\beta }\rightarrow
\infty $, the above entangled coherent states become Bell states which can
be employed to complete teleportation of continuous variables[47]. In
addition, in the initial state of eq.(9), if we let $\left| \psi
\right\rangle _{2}=\left| g\right\rangle _{2}$ and $\left| \psi
\right\rangle _{c,r}=N_{\pm }[\left| \alpha \right\rangle _{c}\pm \left|
-\alpha \right\rangle _{c}]\otimes \left| \beta \right\rangle _{r}$ (the
internal state $\left| \psi \right\rangle _{1}$ of ion $1$ may be an
arbitrary superposition of $\left| g\right\rangle _{1}$ and $\left|
e\right\rangle _{1}$), after performing the transformations of eqs.(8) and
(12) successively, i.e. $U_{3}(t_{3})U_{2}(t_{2})$ with $t_{2}=\pi /(4\Omega
\eta \eta _{r})$ and $t_{3}=2\pi /\chi $, we then obtain another type of
entangled coherent states (unnormalized) for motional states

\begin{equation}
\left| \varepsilon _{-}\right\rangle _{c}\left| \varepsilon
_{+}\right\rangle _{r}\pm \left| \varepsilon _{+}\right\rangle _{c}\left|
\varepsilon _{-}\right\rangle _{r}
\end{equation}
with $\varepsilon _{\pm }=(\beta \pm \alpha )/\sqrt{2}$ while the internal
states of the two ions are immune. The odd two-mode coherent states in eqs.
(22) and (23) are maximally entangled states for two-motional modes.

In the generation of the motional entangled coherent states described above,
we used the odd and even coherent states as the input states which can be
prepared easily by the interaction of a ion with lasers [22-23, 26-27]. The
advantages to produce motional states in terms of beam splitter or/and phase
shifter are the simplicity and practicability: it can produces very
complicated nonclassical motional states through simple networks constituted
by them.

{\bf 4. Mach-Zehnder interferometer for the motional states}

The Mach-Zehnder interferometer is a fundamental interferential device which
may be applied to the detection of phase sensitivity. In general, the input
states of the interferometer are Bose modes which creation and annihilation
operators $d^{+}$ and $d$ satisfy: $d^{+}\left| n\right\rangle =\sqrt{n+1}%
\left| n+1\right\rangle $ and $d\left| n\right\rangle =\sqrt{n}\left|
n-1\right\rangle $ with $\left| n\right\rangle $ a number state of the
considered Bose mode. Leibfried, et al.[46] have recently simulated a
Mach-Zehnder interferential experiment in trapped ion system which used the
internal and external degrees of freedom of a trapped ion as the two input
Bose modes of the interferometer respectively. The harmonic oscillator mode
of the c.m. motion of the ion along one axis in the trap served as one of
the Bose modes. In the restricted Hilbert space formed by the two
ground-state hyperfine levels $\left| F=2,m_{F}=-2\right\rangle \equiv
\left| 1\right\rangle $ and $\left| F=1,m_{F}=-1\right\rangle \equiv \left|
0\right\rangle $ of the ion, the raising and lowering operators $\sigma _{+}$
and $\sigma _{-}$ between the levels have the same action as the
annihilation and creation operators $d$ and $d^{+}$ of a Bose field
respectively. Thus the internal spin motion of the trapped ion can simulate
another Bose mode in a restricted two-dimensional Hilbert space. Here we
implement a Mach-Zehnder interferometer by fully employing two vibrational
modes of trapped ions---the c.m. and breathing modes. It need not restrict
one of the input states of the interferometer within the two-dimensional
number Hilbert space and thus can make more general  interferential
experiments. The sketch for this interferometer is depicted in Fig.1. It
consists of two 50/50 vibrational beam splitters $S_{1}$, $S_{2}$ and a
phase shifter $\varphi $ on c.m. mode. The output phonon-number
distributions for the two modes are detected by $D_{1}$ and $D_{2}$. Note
that the Mach-Zehnder interferometer here works in a temporal rather than a
spatial regime. The c.m. and breathing vibrational motions take place in the
same spatial region which don't propagate along separate paths in practice,
and the introduction of the mirrors in Fig.1 is only for the convenience of
illustrating which don't exist actually. In the following, we describe
concretely the implement of each element. Now the initial state of the whole
system is prepared as

\begin{equation}
\left| \psi _{0}\right\rangle =\left| \psi _{0}\right\rangle _{1}\otimes
\left| g\right\rangle _{2}\otimes \left| in\right\rangle
\end{equation}
with $\left| \psi _{0}\right\rangle _{1}=(\left| e\right\rangle _{1}-\left|
g\right\rangle _{1})/\sqrt{2}$ the internal state of ion $1$ and $\left|
in\right\rangle $ the input motional state of c.m. and breathing modes
entering the interferometer. At the action of eq.(7) for an interaction time 
$T=\pi /(4\Omega \eta \eta _{r})$, the evolved state is

\begin{equation}
\left| \psi _{1}\right\rangle =\left| \psi _{0}\right\rangle _{1}\otimes
\left| g\right\rangle _{2}\otimes e^{i(\pi /2)J_{x}}\left| in\right\rangle
\end{equation}
In this way, we realize the transformation of motional state for the beam
splitter $S_{1}$. By performing the unitary transformation of eqs.(12) on
this state, we then complete the phase shift $\varphi $ on c.m. mode and the
resulted state becomes

\begin{equation}
\left| \psi _{2}\right\rangle =\left| \psi _{0}\right\rangle _{1}\otimes
\left| g\right\rangle _{2}\otimes e^{i\varphi a^{+}a}e^{-i(\pi
/2)J_{x}}\left| in\right\rangle
\end{equation}
with $\varphi =\eta ^{2}\Omega ^{2}t/(4\Delta )$. Performing again the
propagator of eq.(7) for the same interaction time $T$, the final
transformation of beam splitter $S_{2}$ is completed and the output state
becomes as

\begin{equation}
\left| \psi _{3}\right\rangle =\left| \psi _{0}\right\rangle _{1}\otimes
\left| g\right\rangle _{2}\otimes \left| out\right\rangle ,
\end{equation}
with 
\begin{equation}
\left| out\right\rangle =e^{i(\pi /2)J_{x}}e^{i\varphi a^{+}a}e^{i(\pi
/2)J_{x}}\left| in\right\rangle
\end{equation}
the output two-mode motional state. According to the idea of Mach-Zehnder
interferometer, we should measure the phonon-number difference in the two
modes at the output of the interferometer, or equivalently, the mean value: $%
\left\langle J_{z}\right\rangle =\left\langle out\right| J_{z}\left|
out\right\rangle $. Assume that the input motional state is

\begin{equation}
\left| in\right\rangle =\left| 0\right\rangle _{c}\otimes \left| \alpha
\right\rangle _{r},
\end{equation}
i.e. the c.m. mode be in a vacuum state $\left| 0\right\rangle _{c}$ and the
breathing mode in a coherent state $\left| \alpha \right\rangle _{r}$%
(usually with large amplitude $|\alpha |$), then we have

\begin{equation}
\left\langle J_{z}\right\rangle =\frac{1}{2}n\cos \varphi
\end{equation}
with $n=|\alpha |^{2}$ the initial phonon number of breathing mode. It is
shown obviously that if the mean value of $\left\langle J_{z}\right\rangle $
can be measured by experiment, then the shifted phase $\varphi $ may be
determined. The phase uncertainty for this detection may be estimated. From
eqs.(28)-(29), we have

\begin{eqnarray}
\left\langle J_{z}^{2}\right\rangle &=&\left\langle out\right|
J_{z}^{2}\left| out\right\rangle  \nonumber \\
&=&\frac{1}{4}n(1+n\cos ^{2}\varphi ),
\end{eqnarray}
and hence 
\begin{equation}
(\Delta J_{z})^{2}=\left\langle J_{z}^{2}\right\rangle -\left\langle
J_{z}\right\rangle ^{2}=\frac{1}{4}n.
\end{equation}
The phase error is thus given by

\begin{equation}
\Delta \varphi =\frac{\Delta J_{z}}{|\partial \left\langle
J_{z}\right\rangle /\partial \varphi |}=\frac{1}{\sqrt{n}\sin \varphi }.
\end{equation}
When $\varphi =\pi /2$, the phase error is limited to the minimum $1/\sqrt{n}
$ which is reverse to the square root of the input phonon number. Of course,
the sensitivity of an interferometer can be improved by designing the input
states appropriately [1].

The function of the detectors $D_{1}$ and $D_{2}$ is to measure the phonon
number difference of the output two motional modes, i.e. $\left\langle
J_{z}\right\rangle $. In trapped ion system, the measurement of motional
states is usually be realized by first mapping these external degrees of
freedom onto the internals and then making a measurement on the internal
states. The measurement of internal states may be finished by the method of
electron shelving which has the detection efficiency nearly close to
one[48]. Below, we present three methods to measure $\left\langle
J_{z}\right\rangle $. The first way is to measure the phonon number
distributions of the two output modes independently, which may be performed
in principle by use of the dynamics of Jaynes-Cummings model [34, 39]. A
more general measurements of motional states including the determination of
non-diagonal elements of a density matrix may be realized by the method of
quantum-state reconstruction [37]. In the above two-ion system, if we let
ion $2$ interact with a standing wave with the ion located in the node of
the standing wave, we can then get the Hamiltonian as

\begin{equation}
H=\omega _{0}\sigma _{z2}/2+\mu a^{+}a+\nu b^{+}b+H_{int},
\end{equation}
where $H_{int}=(\Omega /2)\sigma _{+2}\exp (-i\omega _{L}t)\sin (k\widehat{x}%
_{2})+h.c.$ with $k$ ($\omega _{L}$) the effective wavevector (frequency) of
the standing wave and $\widehat{x}_{2}$ ($\sigma _{+2}$) the position
operator (internal raising operator) of ion $2$. The meanings of the other
symbols are defined as before. By adjusting the laser frequency to be
resonant with the first red sideband of the c.m. mode, i.e. $\omega
_{L}=\omega _{0}-\mu $, in Lamb-Dicke limit and under rotating wave
approximation, the interaction Hamiltonian in the interaction picture can be
reduced as the standard JCM

\begin{equation}
H_{int}=\lambda (\sigma _{+2}a+\sigma _{-2}a^{+}).
\end{equation}
with $\lambda =\eta \Omega /2$ the coupling strength. For convenience of
narration, we expand the output motional state of eq.(28) as

\begin{equation}
\left| out\right\rangle =\sum_{m,n}c_{mn}\left| m\right\rangle _{c}\left|
n\right\rangle _{r}
\end{equation}
with $\left| m\right\rangle _{c}$, $\left| n\right\rangle _{r}$ the motional
Fock states of c.m. and breathing modes respectively and $c_{mn}$ the
complex numbers. In general, $\left| out\right\rangle $ is a two-mode
entangled state. Performing the JCM interaction of eq.(35) on the
interferometer output state of eq.(27) for a interaction time $\tau $,
followed by a measurement of the internal state of ion $2$, then the
probability that the ion $2$ in the ground state can be expressed as

\begin{equation}
P_{g}(\tau )=\frac{1}{2}[1+\sum_{m}p_{m}\cos (2\lambda \tau \sqrt{m})]
\end{equation}
with $p_{m}=\sum_{n}|c_{mn}|^{2}$ the phonon number distribution of c.m.
mode. A Fourier transformation of the measured signal $P_{g}(\tau )$ allows
one to determine the distribution probability $p_{m}$ of the output c.m.
mode[34, 39]. The phonon-number distribution probability $%
p_{n}=\sum_{m}|c_{mn}|^{2}$ for breathing mode can be obtained in a similar
way. After obtained the phonon-number distribution probabilities $p_{m}$ and 
$p_{n}$ of the two output modes, we can then easily determine the mean value
of $J_{z}$ as 
\begin{equation}
\left\langle J_{z}\right\rangle =\frac{1}{2}(\sum_{m}mp_{m}-\sum_{n}np_{n}).
\end{equation}

The second way to determine $\left\langle J_{z}\right\rangle $ is to measure
the expansion coefficient $c_{mn}$, or equivalently the joint distribution
probability $p_{mn}=|c_{mn}|^{2}$. This can be accomplished in terms of the
dynamics of two-modes JCM. In the above laser-ion interaction, if we let the
ion $2$ locate at the anti-node instead of node of the standing wave, and
adjust the frequency of the standing wave to excite simultaneously the both
lower sidebands of the two motional modes, i.e. $\omega _{L}=\omega
_{0}-(\mu +\nu )$, in Lamb-Dicke limit and under rotating wave
approximation, the interaction Hamiltonian in interaction picture is

\begin{equation}
H_{int}^{\prime }=g(ab\sigma _{+2}+a^{+}b^{+}\sigma _{-2})
\end{equation}
with $g=\Omega \eta \eta _{r}$ the coupling strength. This kind of two-mode
JCM interaction has been studied extensively in quantum optics [49]. Now if
we let eq.(27) as the initial state of this JCM, then the dynamical
evolution, after inserting eq.(36), leads to

\[
\left| \psi (t)\right\rangle =\sum_{m,n}c_{mn}[\cos (gt\sqrt{mn})\left|
m,n,g\right\rangle -i\sin (gt\sqrt{mn})\left| m-1,n-1,e\right\rangle ], 
\]
where the first, second and third marks in Dirac symbols denote the states
of c.m. mode, breathing mode and the internal state of ion $2$ respectively.
The unchanged internal state of ion $1$ is eliminated. Now, if we again
measure the internal state of ion $2$, then the measured probability in
ground state is

\begin{equation}
P_{g}(t)=\frac{1}{2}[1+\sum_{m,n}p_{mn}\cos (2gt\sqrt{mn})].
\end{equation}
In a similar way, we can determine the joint probability distribution $%
p_{mn} $ of the motional state $\left| m,n\right\rangle _{c,r}$ by Fourier
transforming the measured signal $P_{g}(t)$ and then obtain the mean value
of $J_{z}$ in terms of eq.(38) as well as $p_{m}=\sum_{n}p_{mn}$ and $%
p_{n}=\sum_{m}p_{mn}$.

The third way, which is also the simplest way, to determine $\left\langle
J_{z}\right\rangle $ is the direct measurement of the mean value of a
motional operator[40]. In order to do this, we first perform a $\pi /2$%
-pulse carrier transition[48] on ion $2$ to reset the output state of the
interferometer of eq.(27) as

\begin{equation}
\left| \psi _{4}\right\rangle =\left| \psi _{0}\right\rangle _{1}\otimes
\left| \psi _{0}\right\rangle _{2}\otimes \left| out\right\rangle
\end{equation}
with $\left| \psi _{0}\right\rangle _{2}=(\left| g\right\rangle _{2}+i\left|
e\right\rangle _{2})/\sqrt{2}$ the internal state of ion $2$. Then we
execute the unitary transformation $U_{3}$ of eq.(12) on this state. At this
moment, the mean value of $\left\langle \sigma _{x2}\right\rangle $ in the
final state is

\begin{eqnarray}
\left\langle \sigma _{x2}\right\rangle &=&\left\langle \psi _{4}\right|
U_{3}^{+}\sigma _{x2}U_{3}\left| \psi _{4}\right\rangle  \nonumber \\
&=&-\left\langle out\right| \sin (2\chi ta^{+}a)\left| out\right\rangle .
\end{eqnarray}
For small interaction time and not very large number of phonons, we can
linearize the operator as[40] $\sin (2\chi ta^{+}a)\simeq 2\chi ta^{+}a$,
then the above equation reduces to

\begin{equation}
\left\langle \sigma _{x2}\right\rangle =-2\chi t\left\langle out\right|
a^{+}a\left| out\right\rangle .
\end{equation}
In trapped ion system, the measurement of $\left\langle \sigma
_{x2}\right\rangle $ is very simple which may be completed by applying a
single qubit rotation to transform the internal state of ion $2$ from $%
\sigma _{z}$-picture to $\sigma _{x}$-picture, followed by a detection of
the ion's fluorescence[48]. Therefore the mean number of phonons of the c.m.
mode in the output state of the interferometer $\left\langle
a^{+}a\right\rangle $ can be evaluated easily from eq.(43). The mean number
of phonons of breathing mode $\left\langle b^{+}b\right\rangle $ can also be
evaluated in a similar way. Thus the mean value of $\left\langle
J_{z}\right\rangle =[$ $\left\langle a^{+}a\right\rangle -\left\langle
b^{+}b\right\rangle ]/2$ could be obtained.

{\bf 5. Conclusion}

In conclusion, we have proposed a method to realize the transformations of
beam splitter and phase shifter in the motion of trapped ions and discussed
their applications in the generation of quantum motional states and in the
Mach-Zehnder interferometer. Several detection methods for the output
motional state of the interferometer have also been discussed. Further, our
research works may be directly applied to implement some other
interferential experiments of motional states, such as the Homodyne
detection and Michelson-type interferometer [10]. Though the implementation
process involves some complicated ion-laser interactions, some pivotal
techniques such as the basic manipulations of ion-laser interaction
[34-36,43,48] and the reconstruction of motional states[37] have been
already realized experimentally. We expect that, with the increasing of the
experimental precision, our proposals will become truth and produce an
important effects in the future coherent control of quantum motional states
of trapped ions.

{\bf Acknowledgments}

This work was supported by the National Fundamental Research Program Grant
No. 2001CB309310, the National Natural Science Foundation of China under
Grant Nos.10347128, 10325523, 90203018 and 19734006, the EYTF of the
Education Ministry of China, STF of Hunan Province, and the Scientific
Research Fund of Hunan Provincial Education Department.

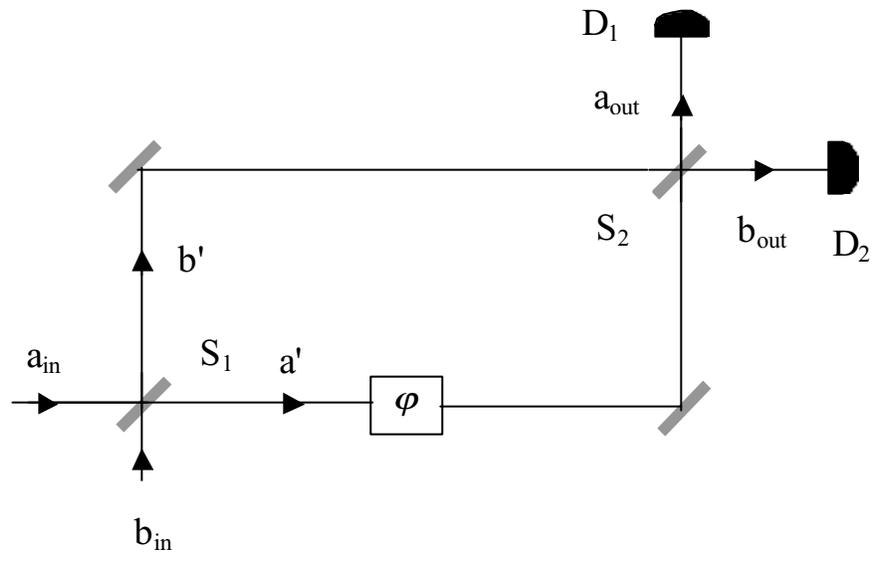
\end{document}